# Human-Centred AI in FinTech: Developing a User Experience (UX) Research Point of View (PoV) Playbook

Festus Adedoyin

Department of Computing and Informatics, Bournemouth University Poole, UK, fadedoyin@bournemouth.ac.uk

Huseyin Dogan

Department of Computing and Informatics, Bournemouth University Poole, UK, hdogan@bournemouth.ac.uk

Advancements in Artificial Intelligence (AI) have significantly transformed the financial industry, enabling the development of more personalised and adaptable financial products and services. This research paper explores various instances where Human-Centred AI (HCAI) has facilitated these advancements, drawing from contemporary studies and industry progress. The paper examines how the application of HCAI-powered data analytics, machine learning, and natural language processing enables financial institutions to gain a deeper understanding of their customers' unique needs, preferences, and behavioural patterns. This, in turn, allows for the creation of tailored financial solutions that address individual consumer requirements, ultimately enhancing overall user experience and satisfaction. Additionally, the study highlights the integration of AI-powered robo-advisory services, which offer customised investment recommendations and portfolio management tailored to diverse risk profiles and investment goals. Moreover, the paper underscores the role of AI in strengthening fraud detection, risk assessment, and regulatory compliance, leading to a more secure and adaptable financial landscape. The findings of this research demonstrate the substantial impact of Human-Centred AI on the financial industry, offering a strategic framework for financial institutions to leverage these technologies. By incorporating a User Experience Research (UXR) Point of View (PoV), financial institutions can ensure that AI-driven solutions align with user needs and business objectives.

CCS CONCEPTS • Human-centred computing • Human and societal aspects of security and privacy • Human computer interaction

**Additional Keywords and Phrases:** Human-Centred AI, Personalised Finance, Adaptable Financial Services, Data Analytics, Machine Learning, Robo-Advisor, Fraud Detection, Risk Assessment, Regulatory Compliance, FinTech UX Research

## 1 INTRODUCTION

The integration of Artificial Intelligence (AI) technologies within the financial technology (FinTech) sector, with a primary focus on enhancing user experience and prioritising consumer needs and preferences, is referred to as Human-Centred AI (HCAI) in FinTech. As the FinTech industry rapidly evolves, the widespread adoption of AI technologies has become increasingly prevalent. HCAI in FinTech now extends beyond the mere application of technology and underscores the importance of prioritising user experience. Consequently, FinTech companies are deploying AI to customise their services and cater to the specific requirements of users, ultimately leading to enhanced customer satisfaction and loyalty.

Consequently, Human-Centred AI has enabled financial institutions to move beyond traditional, rule-based decision-making towards data-driven, adaptive financial services. However, despite AI's potential, many financial organisations struggle to communicate research insights in a way that effectively influences product design and stakeholder decision-making. This paper introduces a UXR playbook tailored for FinTech professionals, providing a structured approach to developing, establishing, and implementing a compelling research PoV that aligns AI-driven insights with user expectations and regulatory considerations.

The next section provides some related work and is followed by section 3 which proposes some point of view building blocks, and playbooks for use within the FinTech space when integrating AI. Section 4 discusses the process as well as ethical and regulatory limitations that should be considered when implementing the plays in the UXR playbook. The paper concludes in section 5 with an agenda for future research.

## 2 RELATED WORK

Research on AI in FinTech has emphasised the importance of data-driven decision-making, predictive analytics, and automation in improving financial services. Arner et al. (2017) discuss how FinTech has reshaped global financial services through innovations in regulation and digital banking. Their work highlights the increasing role of AI-driven automation in risk assessment and compliance, reducing reliance on human analysts while enhancing regulatory transparency. Similarly, Finlay (2018) explores machine learning's impact on business applications, illustrating how AI-driven personalisation in financial services enables institutions to predict customer needs with greater accuracy. This work underlines the importance of integrating AI-powered insights into strategic decision-making to ensure financial products align with evolving consumer behaviour.

Gomber et al. (2018) examine the role of digital finance and predictive modelling in shaping customer experiences. Their study suggests that AI-driven analytics, particularly in fraud detection and real-time financial decision-making, is key to enhancing trust and security in FinTech. Additionally, Brummer & Yadav (2019) delve into the 'innovation trilemma' in FinTech, arguing that technological advancements must balance financial stability, market efficiency, and consumer protection. Their work supports the notion that AI-driven solutions should be designed with a strong ethical foundation to mitigate biases and enhance explainability. Cowgill & Tucker (2020) investigate algorithmic bias and fairness in AI applications, highlighting the risks of automated financial decision-making systems. Their research reinforces the need for a user-centred approach to AI governance in FinTech, ensuring equitable outcomes for all user demographics.

Dogan et al. (2024) present a UXR Point of View (PoV) playbook that provides a set of criteria, standards, and tools to guide multi-disciplinary teams in designing services. Giff et al. (2024) states that a playbook typically consists of a book of plays or a set of instructions and has been used by practitioners from multiple disciplines. This paper utilises the existing UXR PoV to propose an HCAI FinTech UXR playbook as an exemplar case study.

## 3 PROPOSED CONTRIBUTION: THE HCAI FINTECH UXR PLAYBOOK

### 3.1 PoV Building Blocks and Analogical Thinking Accessibility

The HCAI FinTech UXR playbook is structured around key PoV building blocks, enabling financial institutions and UX researchers to craft a compelling research perspective. These building blocks include:

- **Defining the User Context** – Understanding the primary user needs, expectations, and challenges in financial decision-making. This involves leveraging AI-powered behavioural analysis to segment users based on financial habits and goals (Gomber et al., 2018).

- **Establishing the Core Insight** – Identifying the most critical findings from AI-driven research that impact product development. This includes insights from AI-powered personalisation, risk assessment, and customer support interactions.
- **Developing the Research Narrative** – Structuring findings into an engaging and persuasive story that resonates with stakeholders. This involves bridging AI analytics with financial industry challenges, ensuring that insights are both technically robust and strategically relevant (Brummer & Yadav, 2019).
- **Validating with Data and Evidence** – Supporting the PoV with robust quantitative and qualitative data, including customer behaviour trends, case studies, and industry benchmarks (Arner et al., 2017).
- **Translating Insights into Actionable Recommendations** – Converting research findings into concrete product improvements, policy recommendations, or UX enhancements.

Analogical thinking plays a crucial role in shaping a compelling PoV in FinTech. By drawing parallels between established financial principles and emerging AI-driven trends, researchers can make AI-powered insights more accessible and impactful. For instance, traditional risk assessment models in banking can be compared to machine learning-driven fraud detection, helping stakeholders see AI's value in a familiar context. Similarly, the role of a human financial advisor can be used as an analogy to explain the workings of robo-advisors in investment management.

**3.2 Key Plays**

- **Play 1: Identifying Core User Financial Needs** – Utilising HCAI-powered data analytics and machine learning to identify user pain points and expectations.
    - This includes examining transaction history, spending habits, and digital interactions to tailor financial solutions effectively.
- **Play 2: Framing the AI-Driven Research Narrative** – Structuring AI-generated insights into a clear and persuasive story that aligns with financial stakeholders' interests.
    - By translating complex data models into actionable insights, FinTech firms can ensure AI recommendations resonate with business objectives.
- **Play 3: Aligning with Business and Regulatory Goals** – Ensuring the UXR PoV aligns with financial industry regulations (e.g., GDPR, PSD2) and key business performance indicators.
    - Capturing AI explainability requirements and ethical considerations surrounding financial data use.
- **Play 4: Visualising and Presenting AI-Driven Insights** – Leveraging predictive modelling, data visualisation, and explainable AI (XAI) to enhance transparency and user trust.
    - AI-generated recommendations should be made interpretable for both financial experts and everyday consumers.
- **Play 5: Advocating for Ethical and Transparent AI in Finance** – Ensuring fairness, explainability, and user trust in AI-driven financial products, reducing bias in decision-making.
    - Addressing biases in credit scoring, loan approvals, and automated investment decisions is crucial for fostering consumer confidence.

Each play involves specific tasks aimed at ensuring that the AI in FinTech is designed and deployed with a user-centred approach that addresses both technical functionality and user experience.

**Play 1: Identifying Core User Financial Needs**
**Tasks:**

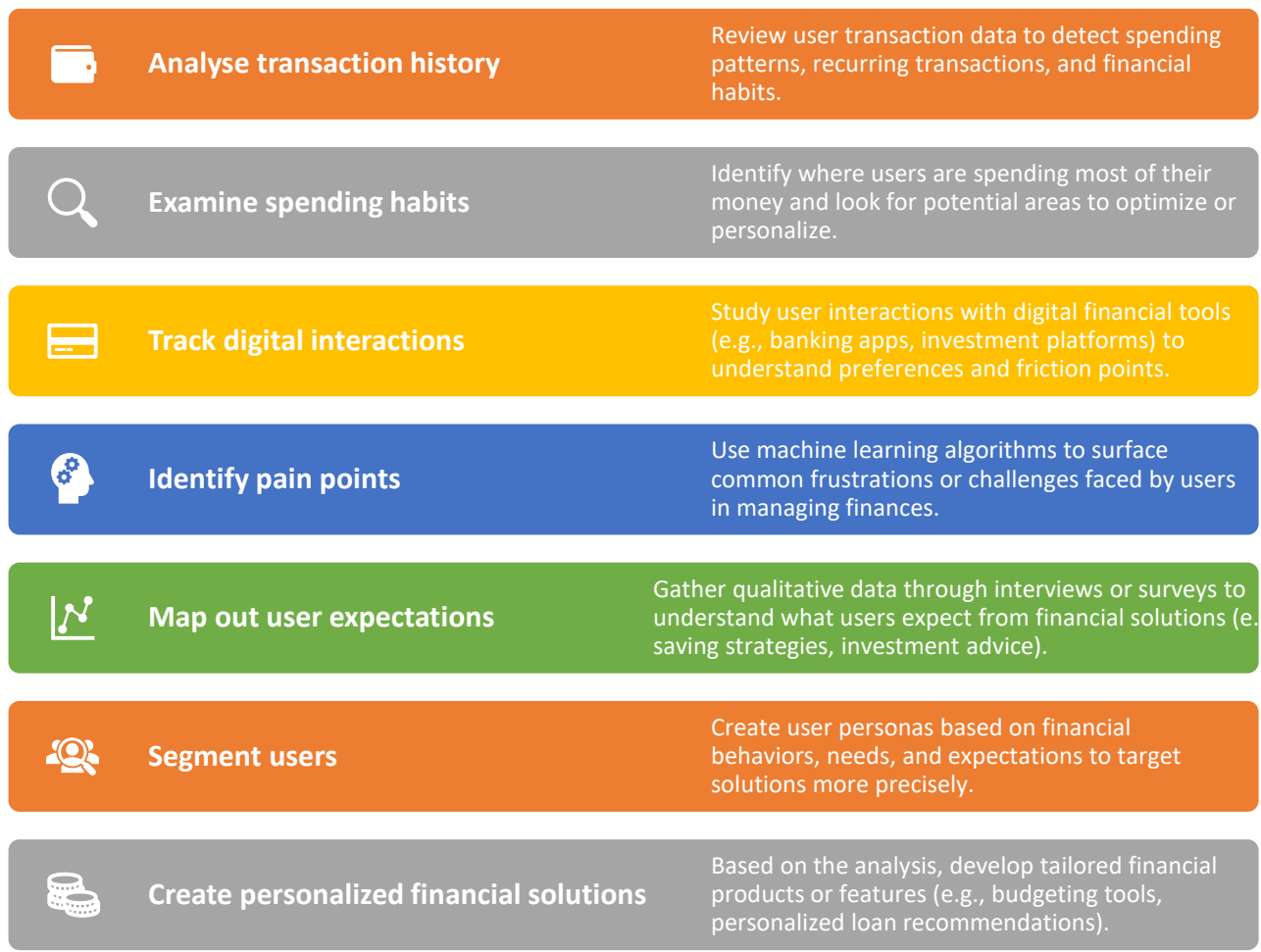

Figure 1: Play 1

**Play 2: Framing the AI-Driven Research Narrative**

**Tasks:**

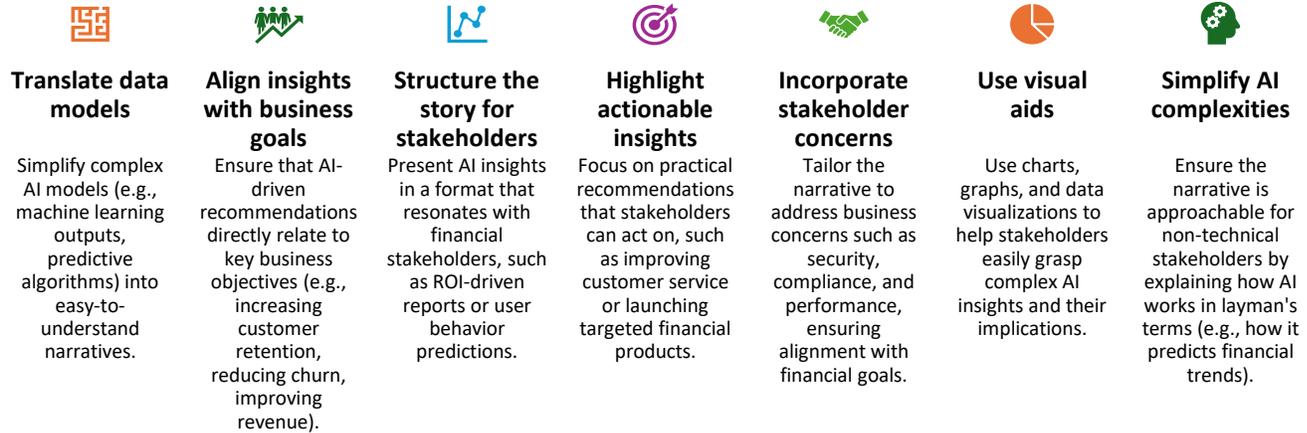

| Translate data models | Align insights with business goals | Structure the story for stakeholders | Highlight actionable insights | Incorporate stakeholder concerns | Use visual aids | Simplify AI complexities |
|---|---|---|---|---|---|---|
| Simplify complex AI models (e.g., machine learning outputs, predictive algorithms) into easy-to-understand narratives. | Ensure that AI-driven recommendations directly relate to key business objectives (e.g., increasing customer retention, reducing churn, improving revenue). | Present AI insights in a format that resonates with financial stakeholders, such as ROI-driven reports or user behavior predictions. | Focus on practical recommendations that stakeholders can act on, such as improving customer service or launching targeted financial products. | Tailor the narrative to address business concerns such as security, compliance, and performance, ensuring alignment with financial goals. | Use charts, graphs, and data visualizations to help stakeholders easily grasp complex AI insights and their implications. | Ensure the narrative is approachable for non-technical stakeholders by explaining how AI works in layman's terms (e.g., how it predicts financial trends). |

Figure 2: Play 2

**Play 3: Aligning with Business and Regulatory Goals**

**Tasks:**

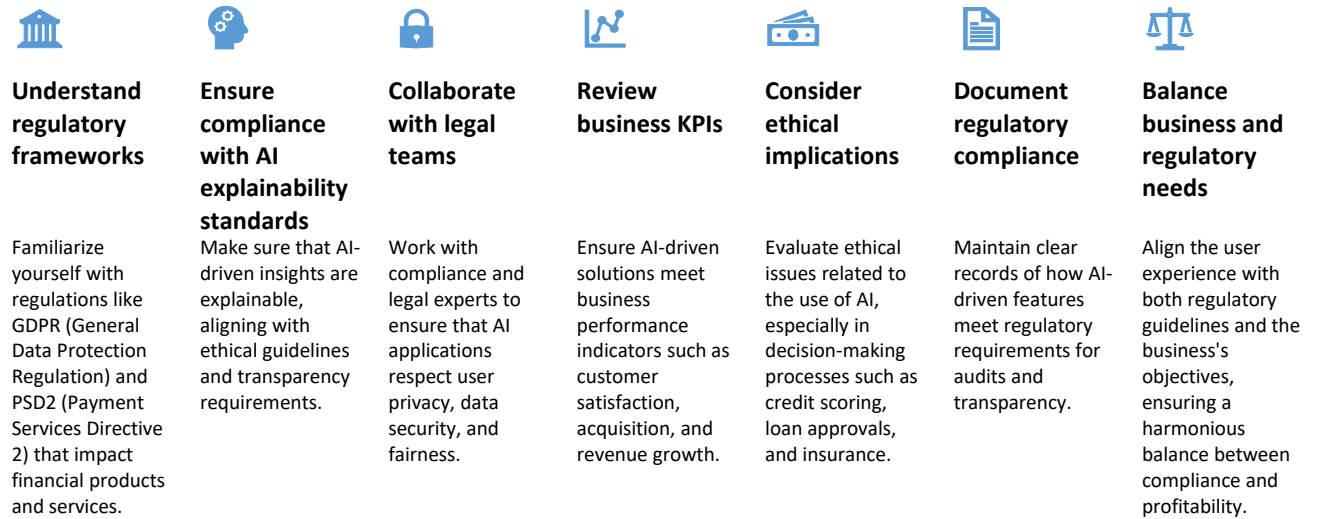

| Understand regulatory frameworks | Ensure compliance with AI explainability standards | Collaborate with legal teams | Review business KPIs | Consider ethical implications | Document regulatory compliance | Balance business and regulatory needs |
|---|---|---|---|---|---|---|
| Familiarize yourself with regulations like GDPR (General Data Protection Regulation) and PSD2 (Payment Services Directive 2) that impact financial products and services. | Make sure that AI-driven insights are explainable, aligning with ethical guidelines and transparency requirements. | Work with compliance and legal experts to ensure that AI applications respect user privacy, data security, and fairness. | Ensure AI-driven solutions meet business performance indicators such as customer satisfaction, acquisition, and revenue growth. | Evaluate ethical issues related to the use of AI, especially in decision-making processes such as credit scoring, loan approvals, and insurance. | Maintain clear records of how AI-driven features meet regulatory requirements for audits and transparency. | Align the user experience with both regulatory guidelines and the business's objectives, ensuring a harmonious balance between compliance and profitability. |

Figure 3: Play 3

**Play 4: Visualising and Presenting AI-Driven Insights**
**Tasks:**

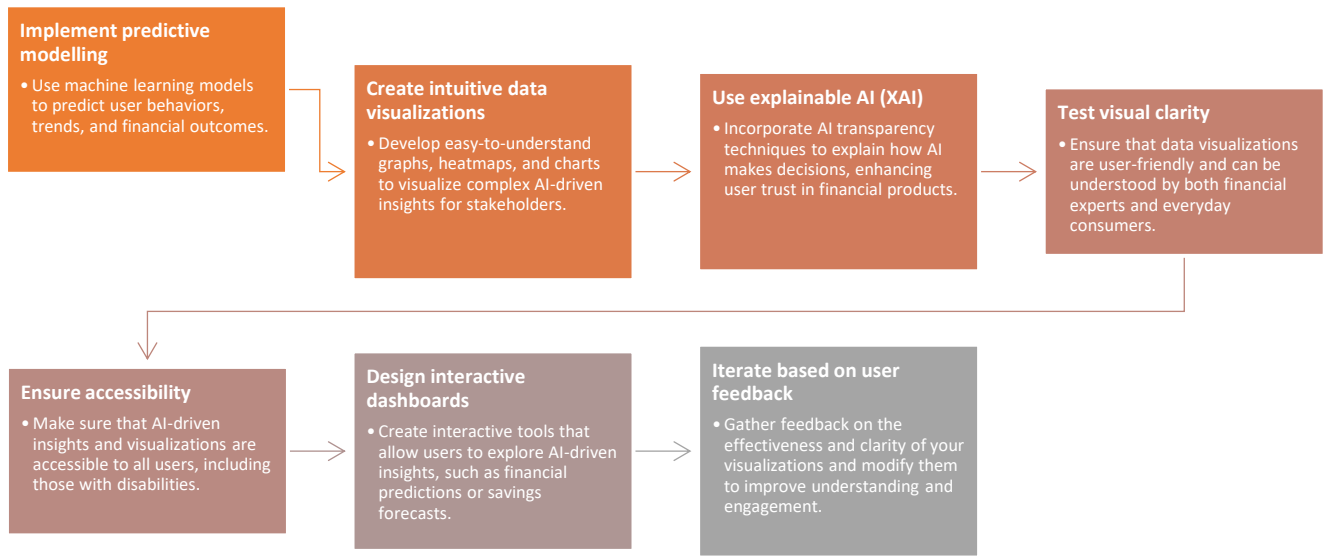

Figure 4: Play 4

**Play 5: Advocating for Ethical and Transparent AI in Finance**

**Tasks:**

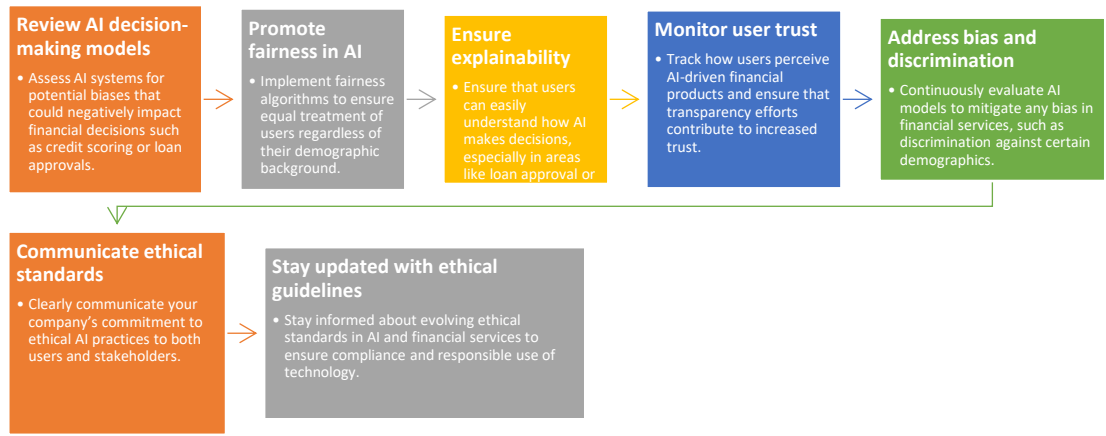

Figure 5: Play 5

These tasks outline how you can carry out each play effectively in a UX research context for AI-driven financial products.

## 4  DISCUSSIONS: IMPLEMENTATION IN THE FINTECH UXR PLAYBOOK

To operationalise these plays, the research should consider providing Templates and Guidelines that includes step-by-step instructions on implementing each play in financial UX research; established case studies that are real-world examples illustrating successful AI-powered UXR PoV articulation in FinTech, such as the adoption of AI-driven risk assessment tools by digital banks and the use of robo-advisors in wealth management; as well as stakeholder engagement strategies which includes methods for ensuring AI-driven research findings influence financial product design, compliance decisions, and business strategies. This includes techniques for communicating AI insights to executives, policymakers, and UX teams.

Additionally, it is vital to understand Human-Centred AI in FinTech by considering not only the development of advanced computational techniques and forecasting models, but also a deep understanding of human behaviour, financial decision-making processes, preferences, needs, and the underlying psychology of money management. By integrating insights from the fields of behavioural economics and cognitive psychology, FinTech companies are designing AI-powered solutions that align with the way users conceptualise and make financial decisions, thereby providing a seamless user experience (Irimia-Diéguez et al. 2023). These sources emphasise the importance of understanding user emotions, goals, and decision-making processes to create AI-driven financial services that are not only efficient but also intuitive and empowering for users. By incorporating these insights into AI-based technologies, financial institutions have transformed the way they engage and interact with their customers.

There is a prominence of AI-Powered Solutions in FinTech especially the AI-powered robo-advisor, which is gaining traction in the market by offering customised investment recommendations and portfolio management services. Research by Pal et al. (2023) and Chen highlights the growing popularity of these AI-powered robo-advisors, as they provide personalised investment guidance and portfolio management solutions. Similarly, Luo et al. (2023) discuss the application of AI in fraud detection, illustrating its ability to identify suspicious patterns and mitigate financial risks in real-time (Ashta and Herrmann 2021). Additionally, the integration of AI into mobile banking and online finance platforms enhances user experience by streamlining processes, providing real-time insights, and offering personalised recommendations. This is achieved through the integration of natural language processing and machine learning algorithms, which personalise and automate customer interactions, ultimately benefiting users and bolstering trust in the financial system. Research has demonstrated how these AI algorithms improve user experience and decision-making processes in HCAI FinTech solutions, showcasing the disruptive potential of AI (Ashta and Herrmann 2021; Kumar and Gupta 2023) as exemplified by the rise of AI-powered robo-advisors. Furthermore, studies have proposed cost-effective AI solutions for portfolio management, achieving acceptance among traditional firms (Wang and Yu 2021).

Another growing development is virtual assistants and cognitive AI which explores the integration of AI-driven virtual assistants in service offerings and has emerged as a key transformation in the FinTech industry. These virtual assistants engage in natural, contextual dialogues with customers, demonstrating the ability to understand inquiries and provide tailored solutions to enhance customer service and engagement. Functioning as "virtual financial coaches," they offer deeper insights and personalised recommendations based on users' financial goals and behavioural patterns. In fact, the increasing prevalence of AI-powered chatbots and virtual assistants, as highlighted in studies such as Kumar and Gupta, enables FinTech organisations to deliver customised client service and support on a continuous basis. By integrating these AI chatbots with cognitive AI capabilities, FinTech solutions can provide personalised financial advice, respond to natural language queries, and guide users through complex financial decision-making processes. As a result, cognitive AI can automate tasks, offer personalised recommendations, and recognise behaviour patterns, ultimately making financial management more accessible and intuitive for users.

The promoting of financial inclusion, therefore, is enhanced by HCAI in FinTech as it enhances accessibility by utilising AI to analyse data and detect patterns (Ashta and Herrmann 2021). This allows FinTech firms to create solutions tailored to underserved populations, granting them access to previously unavailable financial products and services. AI-powered FinTech solutions thus have the capacity to address the specific requirements of diverse user groups, promoting financial literacy and inclusion. These advancements foster a more equitable financial landscape and drive positive societal transformation.

There are several challenges in HCAI implementation that still needs to be addressed for which this UXR Playbook can proffer bespoke solutions for firms in the FinTech space. Existing research has highlighted the potential risks associated with the use of AI in the financial sector, including algorithmic bias, data privacy concerns, and the potential for job displacement (Wang and Yu 2021). As a result, FinTech companies must carefully navigate these challenges and ensure that AI implementation aligns with ethical principles and regulatory frameworks. To address these issues, FinTech companies should adopt a multifaceted approach. The methods should encompass the implementation of cutting-edge technologies, the incorporation of user feedback into design processes, the establishment of robust governance frameworks for data security, and the fostering of continuous collaboration with stakeholders. These regulatory frameworks must be put in place to safeguard users and address potential risks. They should include data protection laws, consumer rights regulations, and financial conduct standards to ensure that AI technologies operate within legal boundaries and maintain consumer confidence. Furthermore, a focus on ethical AI governance should lead to the establishment of legal frameworks that enforce principles of transparency, fairness, and non-discrimination.

These brings the problems of compliance with ethical principles and other regulatory safeguards as the fundamental principles of transparency, accountability, fairness, and explainability are crucial in addressing potential biases and promoting reliability in AI systems (Han et al. 2023; Maple et al. 2023). This ensures that users understand how AI algorithms make decisions and recommendations. By prioritising explainability, FinTech companies can build user confidence, reduce discrimination, and allow individuals to comprehend the rationale behind algorithmic outcomes. Therefore, striking the right balance between innovation and responsible deployment enables the FinTech industry to leverage the capabilities of HCAI to create a more inclusive, transparent, and trustworthy financial ecosystem (Hu 2020; Ashta and Herrmann 2021). Implementing these principles can help achieve the full potential of HCAI in the FinTech industry, ensuring positive impacts on individuals, businesses, and society.

## 5 Conclusion & Future Work

Employing robust data privacy safeguards, such as responsible management and transparent handling of customer data, can enable financial institutions to build trust with both customers and regulators. This, in turn, can promote ethical financial practices. At its core, the integration of Human-Centred Artificial Intelligence in the FinTech industry combines innovative technologies with user-centric approaches to prioritise customer needs while upholding ethical principles, moral values, and legal requirements.

By defining a structured approach to building a UXR PoV for FinTech, we empower UX practitioners and financial institutions to leverage AI responsibly while aligning with user expectations. Future iterations of this playbook will incorporate feedback from the FinTech and UX research communities, refining the plays and expanding best practices.

We invite practitioners to contribute their insights, case studies, and methodologies to further enhance this resource as we develop an agenda for future research and workshops.